\documentclass[review,12pt]{elsarticle}
\usepackage{amssymb}
\usepackage{amsmath}
\usepackage{natbib}
\usepackage{setspace}
\usepackage{makecell} 
\usepackage[colorlinks]{hyperref}
\usepackage{xcolor}
\usepackage{caption,geometry}
\usepackage{setspace}
\usepackage{bm}
\usepackage{lineno}
\usepackage[percent]{overpic}
\journal{The Innovation}
\biboptions{numbers,sort&compress}
\newcommand{\upcitep}[1]{\textsuperscript{\textsuperscript{\cite{#1}}}}
\setcitestyle{open={},close={}}

\begin{document}
	\let\WriteBookmarks\relax
	\def\floatpagepagefraction{1}
	\def\textpagefraction{.001}
	
    \begin{center}
		{\fontsize{17pt}{17pt}\selectfont The first extragalactic ultra-compact X-ray binary : a candidate black hole-white dwarf system} \\[3ex]
        {\fontsize{11pt}{11pt}\selectfont Qian-Qi Ma,$^{1,6}$ Jiachang Zhang,$^{2,3,6}$ Wei-Min Gu,$^{1,*}$ Zhiyuan Li,$^{2,3,*}$ Shan-Shan Weng,$^{4}$ and Tong Bao$^{5}$}
		\begin{spacing}{1.0}
		$^1$\fontsize{10pt}{14.4pt}\selectfont\textit{Department of Astronomy, Xiamen University, Xiamen, Fujian 361005, China}\\
		$^2$ \fontsize{10pt}{14.4pt}\selectfont\textit{School of Astronomy and Space Science, Nanjing University, Nanjing 210046, China}\\
		$^3$ \fontsize{10pt}
        {14.4pt}\selectfont\textit{Key Laboratory of Modern Astronomy and Astrophysics (Nanjing University), Ministry of Education, Nanjing 210046, China}\\
		$^4$ \fontsize{10pt}{14.4pt}\selectfont\textit{Department of Physics and Institute of Theoretical Physics, Nanjing Normal University, Nanjing 210023, China}\\
        $^5$ \fontsize{10pt}
        {14.4pt}\selectfont\textit{INAF -- Osservatorio Astronomico di Brera, Via E. Bianchi 46, Merate 23807, Italy}\\
        $^6$ \fontsize{10pt}
        {14.4pt}\selectfont\textit{These authors contributed equally to this work}
        \fontsize{10pt}
        {14.4pt} \\
        *Correspondence: \href{mailto:guwm@xmu.edu.cn}{guwm@xmu.edu.cn} (Wei-Min Gu); \href{mailto:lizy@nju.edu.cn}{lizy@nju.edu.cn} (Zhiyuan Li)
        \end{spacing}
	\end{center}	
	\noindent\textbf{GRAPHICAL ABSTRACT}	
	\begin{center}
		\includegraphics[width=0.65\linewidth]{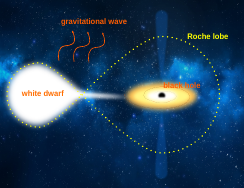}
	\end{center}	
	\noindent\textbf{PUBLIC SUMMARY}
	\begin{itemize}
        \item M31 UCXB-1 is the first extragalactic ultra-compact X-ray binary (UCXB) candidate
		\item M31 UCXB-1 is in excellent agreement with the black hole/neutron star--white dwarf system
		\item The primary in M31 UCXB-1 is more likely to be a black hole rather than a neutron star
        \item This source hosts the shortest orbital period and the most massive white dwarf among the known UCXBs
	\end{itemize}
    \clearpage
	M31 UCXB-1 is one of the brightest X-ray point sources in the bulge of M31, with a peak X-ray luminosity $ L_{\mathrm{0.5-10 \: keV}} = 2.9^{+0.2}_{-0.2} \times 10^{38} \: \mathrm{erg} \: \mathrm{s}^{-1} $. Both \textit{XMM-Newton} and \textit{Chandra} observations have detected an eclipsing signal with a period of about 465 seconds from this source, and we note that the periodic signal is detected exclusively during the source's high-luminosity states. This signal probably originates from its orbital motion, therefore it is an ultra-compact X-ray binary (UCXB) candidate with the highest X-ray luminosity. Our theoretical analyses show that M31 UCXB-1 is in good agreement with the luminosity-orbital period relation ($ L_{\mathrm{2-10 \: keV}}-P_{\mathrm{orb}} $) of the black hole/neutron star--white dwarf (BH/NS--WD) UCXB system. Moreover, our spectral analyses indicate that the primary in M31 UCXB-1 is more likely to be a BH rather than an NS. The results show that M31 UCXB-1 is a BH--WD system, with the shortest orbital period, the possibly strongest gravitational wave emission, and the most massive white dwarf among the known UCXBs.

	\section*{INTRODUCTION}
    \addcontentsline{toc}{section}{INTRODUCTION}
    \phantomsection
    \label{INTRODUCTION}
    Numerous bright X-ray point sources have been identified in M31 by \textit{Einstein} satellite,\upcitep{Trinchieri1991} \textit{XMM-Newton},\upcitep{Stiele2011, De Luca2021} and \textit{Chandra},\upcitep{Kaaret2002, Hofmann2013} the majority of which are X-ray binaries (XRBs).\upcitep{Remillard2006} The X-ray point source at J2000 equatorial coordinates $ 00^{\mathrm{h}} \: 42^{\mathrm{m}} \: 22^{\mathrm{s}}.919 \: + 41^{\mathrm{o}} \: 15' \: 35''.14 $ with the X-ray luminosity $ L_{\mathrm{X}} \sim 10^{38} \: \mathrm{erg} \: \mathrm{s}^{-1} $, is one of the brightest X-ray point sources in M31,\upcitep{Kong2002} which was firstly identified by \textit{Einstein} satellite\upcitep{Trinchieri1991} and Barnard et al.\upcitep{Barnard2011} named it as RX J0042.3+4115. The luminosity of this source is close to the Eddington luminosity of a neutron star (NS), but its X-ray spectrum is harder than that, which suggests that the accretor is more likely to be a black hole (BH).\upcitep{Barnard2011} Recently, several studies have detected a 465 sec periodic signal from this source,\upcitep{De Luca2021, Zhang2024} but the origin of this periodic signal has not yet been determined. Yang et al.\upcitep{Yang2025} suggested that this signal possibly originates from the orbital motion, and their simulation results demonstrate that in some special evolution channels, binary systems can evolve to such a short orbital period.

    If the periodic signal of RX J0042.3+4115 corresponds to the orbital period, then it belongs to the ultra-compact X-ray binary (UCXB) population. Considering that this is the first UCXB candidate in M31, we designate this source as M31 UCXB-1. UCXBs are a sub-population of low-mass X-ray binary (LMXB), which have orbital periods shorter than 80 min,\upcitep{Paczynski1981, Armas Padilla2023} formed by a hydrogen-poor star and an accretor (NS or BH). Depending on the type of companion stars, UCXBs can be separated into two groups: white dwarfs (WDs) as companion stars (BH/NS$-$WD systems), with orbital periods shorter than 40 min; helium stars (He stars) as companion stars, with orbital periods longer than 40 min but shorter than 80 min.\upcitep{Heinke2013, Chen2016, Wang2021} Therefore M31 UCXB-1 belongs to the BH/NS--WD system. A known UCXB, 4U 1728-34, was reported to have an extreme short orbital period of 646 sec,\upcitep{Galloway2010} but it is controversial.\upcitep{Vincentelli2020} We therefore regard 4U 1820-303 as the UCXB with the shortest orbital period. 4U 1820-303 is an NS--WD system with the orbital period of 685 sec,\upcitep{Stella1987} and the peak X-ray luminosity of $ L_{\mathrm{2-10 \: keV}}\sim3.7 \times 10^{37} \mathrm{erg} \: \mathrm{s}^{-1} $.\upcitep{Costantini2012} Compared with 4U 1820-303, M31 UCXB-1 hosts the shorter orbital period ($ \sim 465 $ sec) and the higher persistent X-ray luminosity ($L_{\mathrm{2-10 \: keV}} > 10^{38} \mathrm{erg} \: \mathrm{s}^{-1} $). Therefore, M31 UCXB-1 is the first UCXB with an orbital period shorter than 10 min and the persistent X-ray luminosity higher than $ 10^{38} \: \mathrm{erg} \: \mathrm{s}^{-1} $.

    UCXBs are expected to occupy the same canonical spectral states (low/hard, high/soft, and very‐high states) observed in standard X‐ray binaries, although neutron star UCXBs in the high/soft state exhibit an additional strong boundary‐layer blackbody component rather than a steep power‐law tail.\upcitep{van Haaften2012, Piraino1999} In the low/hard state, black hole XRBs (BHXRBs) exhibit a hard power‐law continuum, reflecting Comptonization in a hot, optically thin corona. The accretion disk is truncated at several tens of gravitational radii and contributes only a weak, cool thermal component.\upcitep{Remillard2006} By contrast, neutron star XRBs (NSXRBs) in their analogous hard state also show Comptonization‐dominated emission but include an additional blackbody‐like component arising from the boundary layer at the stellar surface, so their spectra are generally softer below $\sim 10\ \mathrm{keV}$. In the high/soft state, BHXRBs display a dominant multicolor disk blackbody plus a faint steep power‐law tail ($\Gamma\gtrsim\ 2$). NSXRBs in soft state exhibit both a bright disk component and a strong boundary‐layer blackbody, leading to higher overall luminosities and noticeable curvature in the 3--10 keV band.\upcitep{Mitsuda1984,Church2001} During very‐high (steep power‐law) states, BHXRBs are characterized by a strong nonthermal tail extending beyond 100 keV, while NS sources often trace distinct ``Z" or ``atoll" tracks in color‐color diagrams and display additional emission lines or reflection features from the disk and neutron star surface.\upcitep{Homan2007} Thus, although both classes traverse similar state sequences, the presence of a solid surface and boundary layer in NSXRBs produces unique spectral signatures—particularly a hot blackbody component and altered Comptonization behaviour, which allow them to be distinguished from BHXRBs.
        
	\section*{RESULTS}
        \addcontentsline{toc}{section}{RESULTS}
        \phantomsection
        \label{RESULTS}
    
        \subsection*{Period searching results}
\begin{figure}
        \centering
        \captionsetup{font=normalsize, labelfont=bf}
        \begin{overpic}[width=0.49\linewidth]{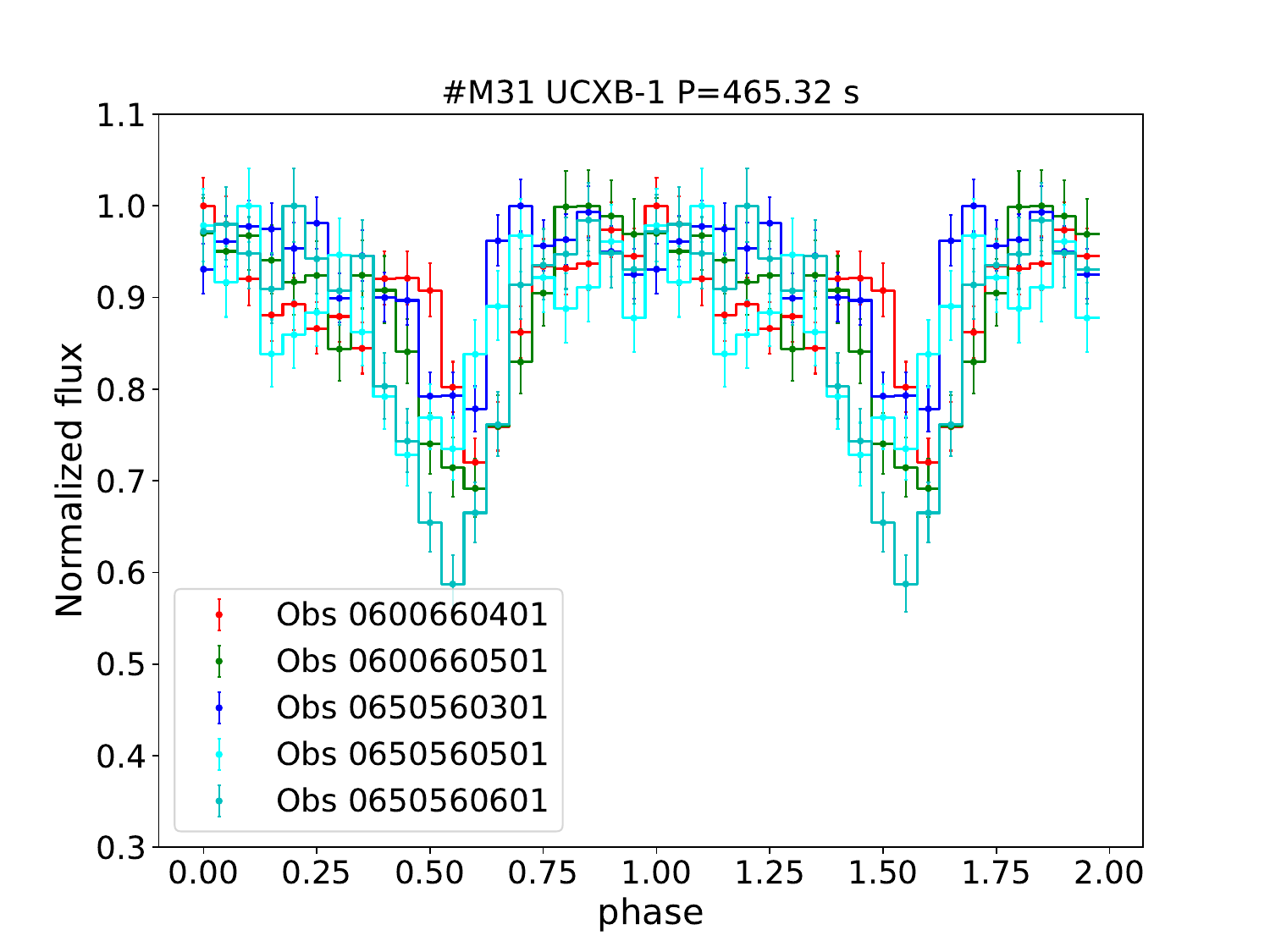}
        \put(0,65){\small A}
        \end{overpic}
        \begin{overpic}[width=0.49\linewidth]{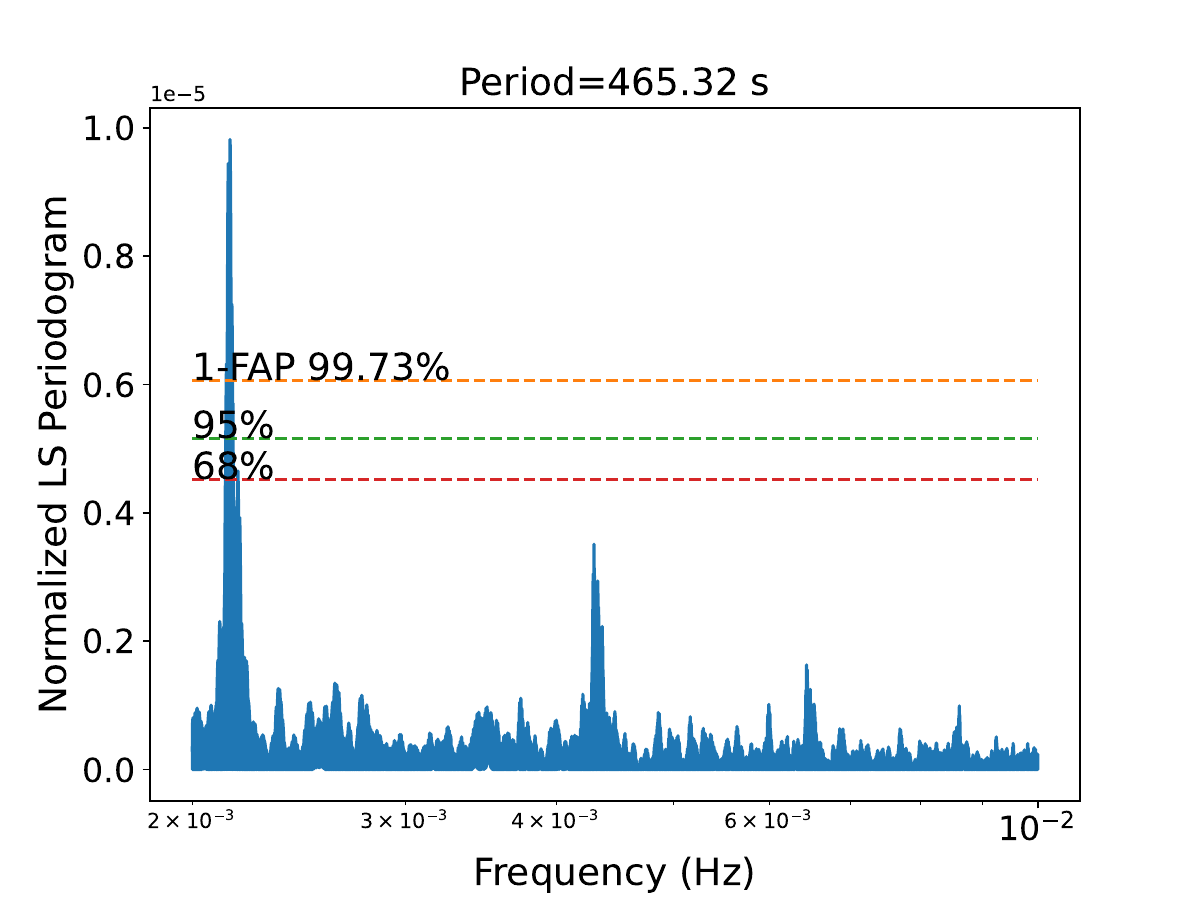}
        \put(0,65){\small B}
        \end{overpic}
        \caption{
        \textbf{The folded light curves of five observations and the normalized Lomb-Scargle periodogram} (A) The five light curves are folded with the period of 465.32 sec. The modulation is stable and the folded profiles from different epochs align in phase, demonstrating the coherence of the signal. (B) The normalized Lomb-Scargle periodogram, displaying a narrow peak at the corresponding frequency with distinct harmonics, confirming the phase stability and orbital nature of the periodicity.}
        \label{fig:1}
    \end{figure}

\begin{figure}
        \centering
        \includegraphics[width=0.7\linewidth]{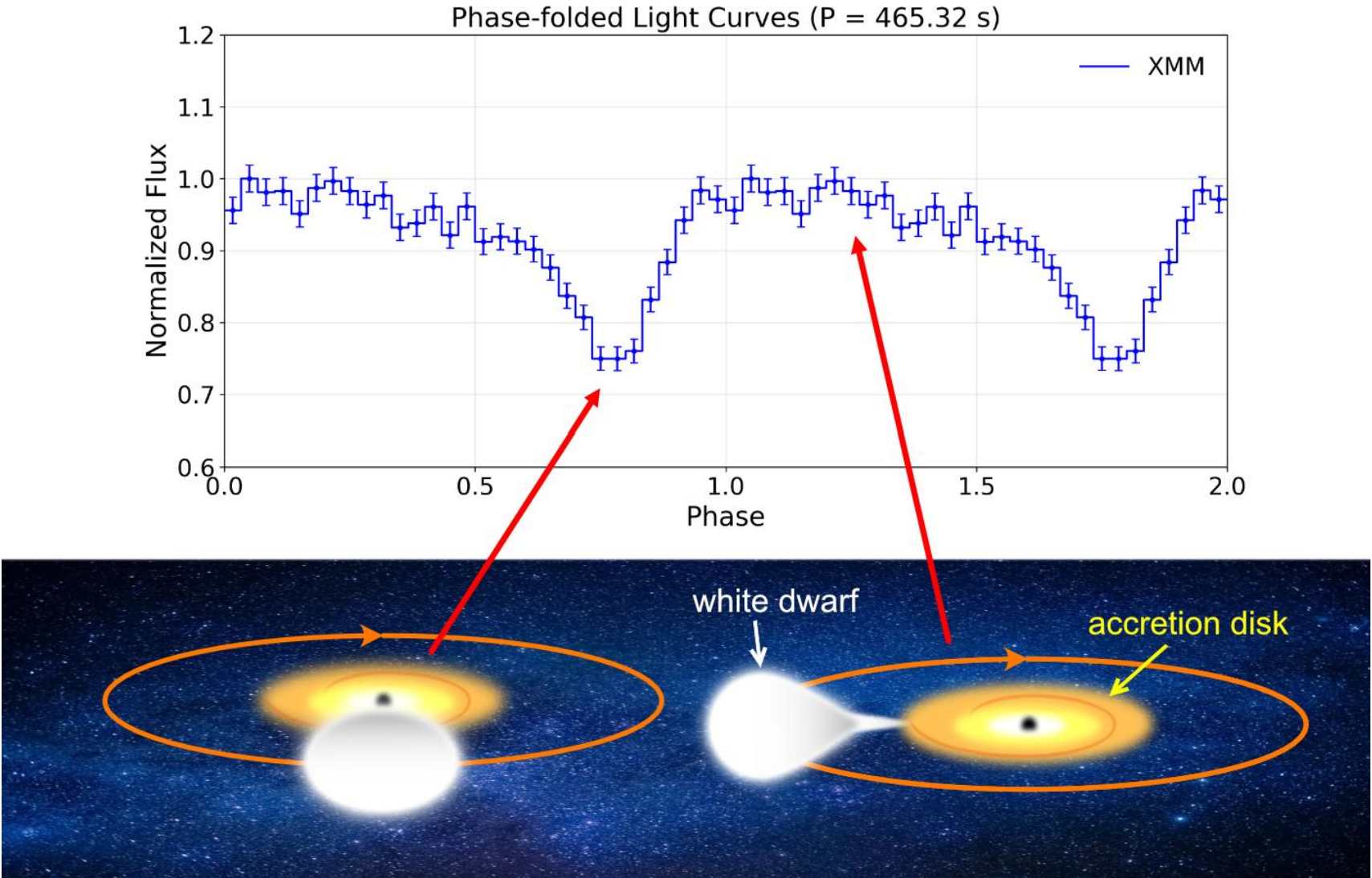}
        \captionsetup{font=normalsize, labelfont=bf}
        \caption{\textbf{The folded light curve from \textit{XMM-Newton} by combining five observations and a schematic picture of the orbital period} The folded light curve exhibits a 465.32 sec period, which originates from the periodic obscuration of the accretion disk by the white dwarf.}
        \label{fig:2}
    \end{figure}

        \addcontentsline{toc}{subsection}{Period searching results}
        \phantomsection
        \label{Period searching results} 
        In our previous work, the \textit{Chandra} observations (ObsIDs 11808, 12113, 12114, 13179, 13180, 11838, 11839, 12164) revealed periodic signal.\upcitep{Zhang2024} No Type-I X-ray bursts were detected in any of the \textit{Chandra} observations. We further analyze the \textit{XMM-Newton} data of M31 UCXB-1 to complement the \textit{Chandra} results.
        Across the 50 \textit{XMM-Newton} observations obtained between 2001 June 29 and 2017 January 21, we detect a coherent modulation in five observations (ObsIDs 0600660401, 0600660501, 0650560301, 0650560501, and 0650560601) with a period of 465.32 sec. An uncertainty $+3.43/-2.99$ sec is derived from the 3$\sigma$ cutoff of the Lomb-Scargle periodogram peak. The normalized Lomb-Scargle periodogram is shown in Figure~\ref{fig:1}, with a narrow peak at 465.32 sec ($2.15\times 10^{-3}$ Hz) and coherent distinct harmonics. This clearly distinguishes M31 UCXB-1 from quasi-periodic oscillations (QPOs), which are known to manifest as broad features in the power spewctrum due to their unstable, non-coherent nature.\upcitep{van der Klis1989,Veresvarska2024,Bao2022} Instead, the presence of a stable, coherent peak with harmonics strongly indicates that the periodicity arises from the binary orbital motion rather than from a QPO phenomenon. Quantitatively, we measure the full width at half maximum (FWHM) of the peak as $\Delta\nu \approx 2.5\times10^{-5}$ Hz, corresponding to a frequency range of 2.135–2.160$\times10^{-3}$ Hz around the centroid at $\nu_0 = 2.15\times10^{-3}$ Hz (465.32 sec). This yields a quality factor $Q = \nu_0/\Delta\nu \sim 86$, far higher than the typical $Q < 10$ observed in stellar-mass black hole QPOs.\upcitep{Belloni2002} Furthermore, the phase-folded light curves from individual observations, also presented in Figure~\ref{fig:1}, align consistently in phase and show repeated modulation at the detected period, demonstrating the phase stability of the signal and providing additional evidence for its orbital origin. We speculate that the periodic X-ray signal originates from the obstruction of the accretion disk by the WD, as shown in Figure \ref{fig:2}.

        \subsection*{Numerical calculation results}
        \addcontentsline{toc}{subsection}{Numerical calculation results}
        \phantomsection
        \label{Numerical calculation results}

        To describe the BH/NS--WD UCXB system, we employ $ M_{1} $ and $ M_{2} $ (or $ M_{\mathrm{WD}} $) as the mass of the accretor and the mass of the donor, respectively. The Roche lobe radius of the companion star is expressed as $ R_{\mathrm{L2}} $, and the orbital separation is expressed as $ a $, with the orbital period $ P_{\mathrm{orb}} $. If the signal with the period of 465.32 sec corresponds to the orbital period, M31 UCXB-1 belongs to the BH/NS--WD UCXB system. In order to verify this, we need to compare this source with the numerical calculation results, and the other UCXBs. Numerical calculations by Gu et al.\upcitep{Gu2020} demonstrate that shorter $ P_{\mathrm{orb}} $ corresponds to smaller $ R_{\mathrm{L2}} $ and higher $ M_{2} $. Following this we infer that M31 UCXB-1 hosts the most massive and the most compact WD compared to other UCXBs. When the WD fills its Roche lobe, $ M_{2} $ depends solely on $ P_{\mathrm{orb}} $ (see the \nameref{BH/NS--WD binary system} section). Combining the WD's $M$-$R$ relation, we obtain $ M_{\mathrm{WD}} = 8.761^{+0.047}_{-0.068} \times 10^{-2} M_{\odot} $ and $ R_{\mathrm{WD}} = 2.462^{+0.006}_{-0.004} \times 10^{-2} R_{\odot} $ for $ P_{\mathrm{orb}} = 465^{+3.43}_{-2.99} \: \mathrm{s} $, where the uncertainties of $ M_{\mathrm{WD}} $ and $ R_{\mathrm{WD}} $ originate from the error in $ P_{\mathrm{orb}} $.

    \begin{figure}
        \centering
        \includegraphics[width=0.7\linewidth]{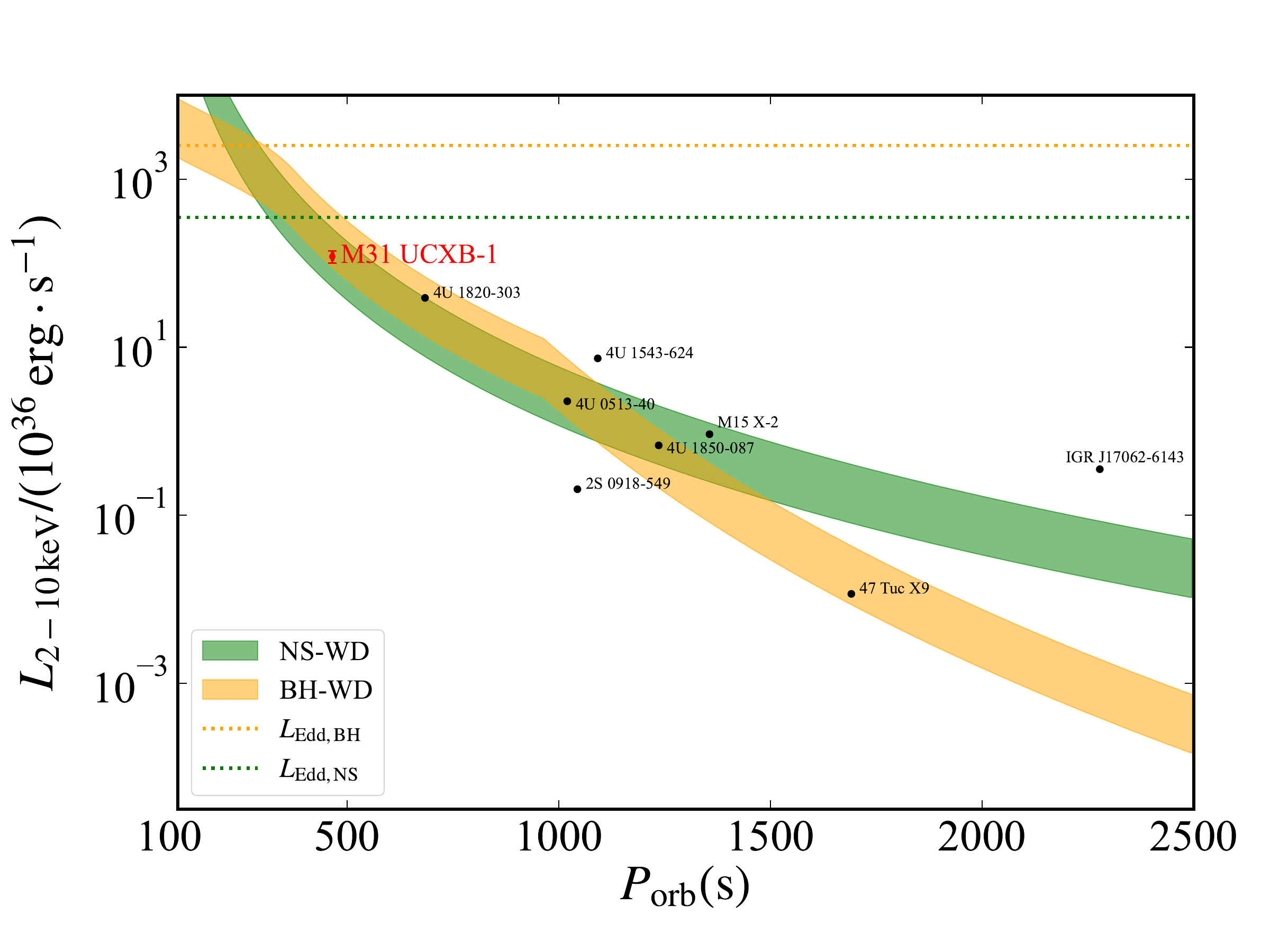}
        \captionsetup{font=normalsize, labelfont=bf}
        \caption{\textbf{The \bm{$ L_{\mathrm{2-10 \: keV}}-P_{\mathrm{orb}} $} diagram} The location of M31 UCXB-1 (red dot, $ L_{\mathrm{2-10 \: keV}} = 1.2^{+0.2}_{-0.2} \times 10^{38} $ erg $ \mathrm{s}^{-1} $, $ P_{\mathrm{orb}} = 465^{+3.43}_{-2.99} \: \mathrm{s} $) and the known UCXBs (black dots) are shown in this figure, with the theoretical predictions for NS--WD (green bands) and BH--WD systems (orange bands). Except for M31 UCXB-1 and 47 Tuc X-9, which are BH candidates, the other known UCXBs in this diagram are all NS systems. The green and orange bands are obtained by assuming that $ M_{\mathrm{BH}} = 10 \: M_{\odot} $ and $ M_{\mathrm{NS}} = 1.4 \: M_{\odot} $, with the bolometric correction factor $ \lambda = L_{\mathrm{2-10 \: keV}} / L_{\mathrm{bol}} \sim 0.1-0.5 $.}
        \label{fig:3}
    \end{figure}	

\begin{figure}
        \centering
        \includegraphics[width=0.7\linewidth]{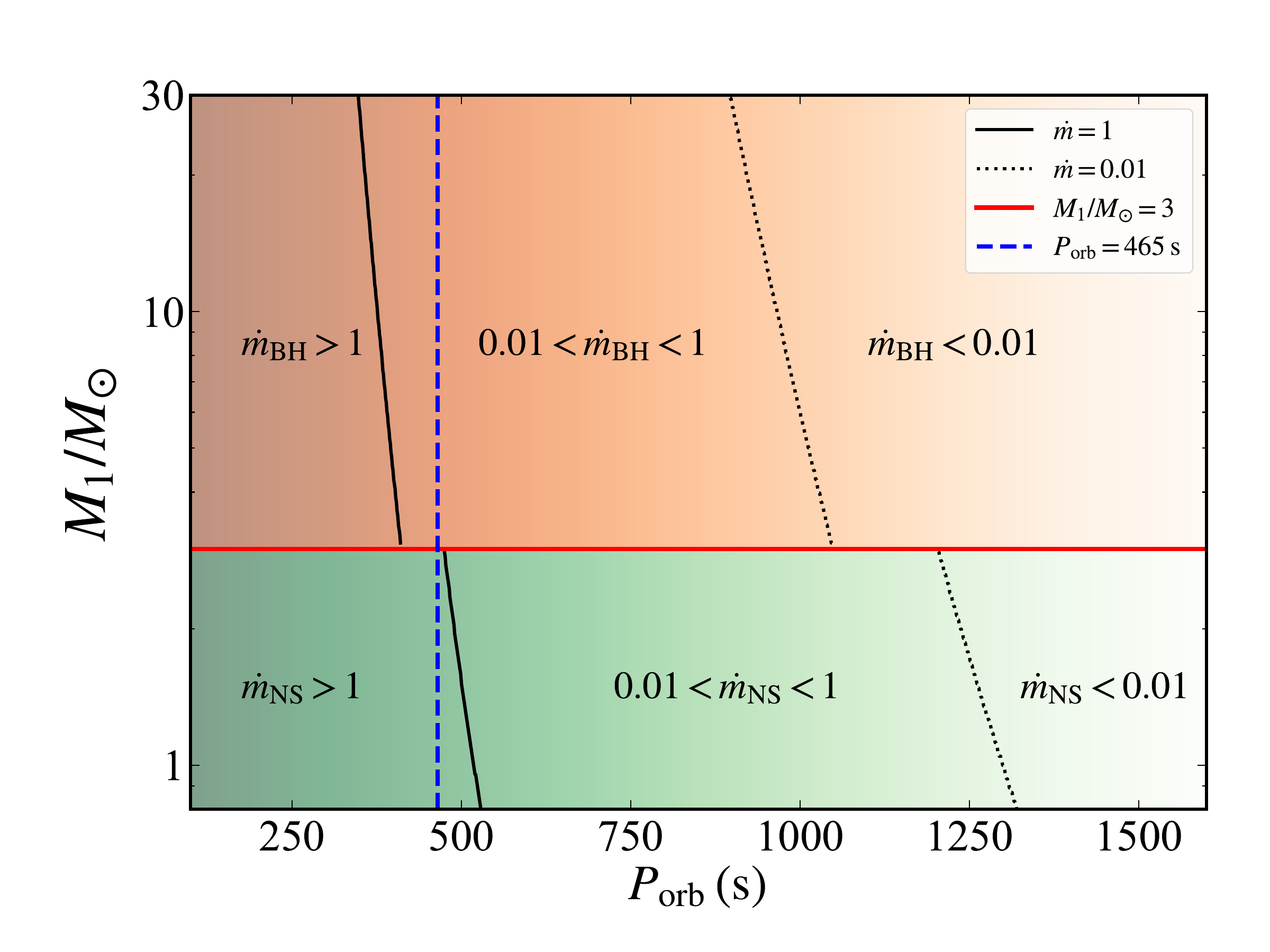}
        \captionsetup{font=normalsize, labelfont=bf}
        \caption{\textbf{The ranges of \bm{$\dot{m}$} as a function of \bm{$ M_{1} / M_{\odot} $} and \bm{$ P_{\mathrm{orb}} $}} The $ M_{1} / M_{\odot} < 3 $ region corresponds to the NS$-$WD systems, while $ M_{1} / M_{\odot} > 3 $ region corresponds to the BH$-$WD systems. The darker region corresponds to higher $ \dot{m} $. For the condition that $ P_{\mathrm{orb}} = 465 \: \mathrm{s} $, the NS$-$WD system is super-Eddington accretion, while for BH$-$WD is sub-Eddington accretion.}
        \label{fig:4}
    \end{figure}

        We cannot determine whether the accretor in M31 UCXB-1 is a BH or an NS solely based on the $ L-P_{\mathrm{orb}} $ diagram, as shown in Figure \ref{fig:3}. The predicted $ L_{\mathrm{2-10 \: keV}} $ for BH and NS systems by the numerical results are similar for $ P_{\mathrm{orb}} < 1000 $ s. However, the accretion states of BH and NS are different when $ P_{\mathrm{orb}} = 465 $ s, which can be compared within the parameter space of the binary model. The two input parameters in our model are $ M_{1} $ and $ P_{\mathrm{orb}} $, and we can divide the parameter space formed by $ M_{1} $ and $ P_{\mathrm{orb}} $ into different regions of accretion rates (see Figure \ref{fig:4}). We define $ \dot{m} = \dot{M}_{1} / \dot{M}_{\mathrm{Edd}} $ as the relative Eddington accretion rate, and Figure \ref{fig:4} exhibits that for an NS$-$WD system, $ P_{\mathrm{orb}} = 465 \: \mathrm{s} $ corresponds to supre-Eddington accretion, while for a BH$-$WD system, $ P_{\mathrm{orb}} = 465 \: \mathrm{s} $ corresponds to sub-Eddington accretion.

\begin{figure}
    \centering
    \includegraphics[width=0.7\linewidth]{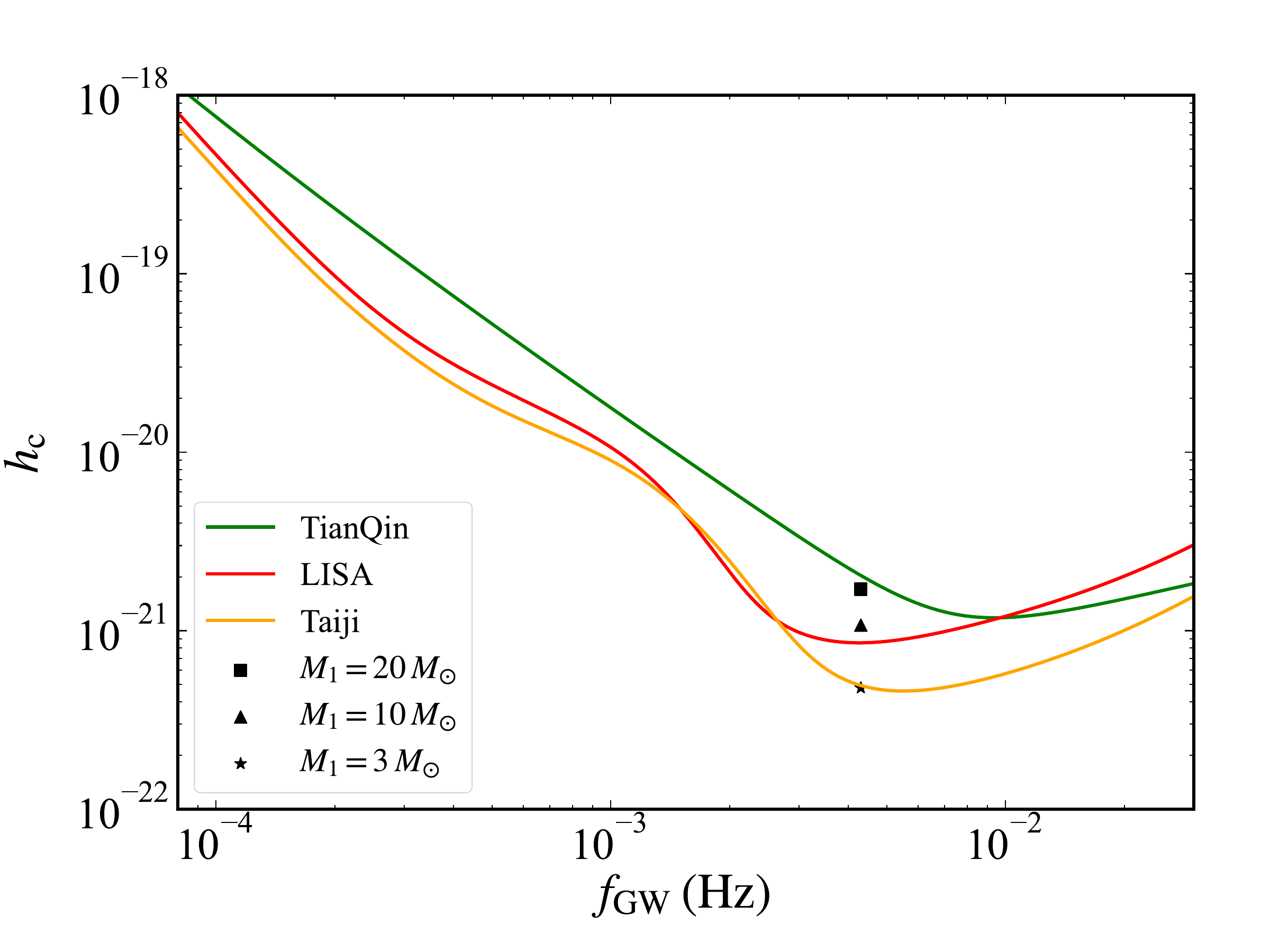}
    \captionsetup{font=normalsize, labelfont=bf}
    \caption{\textbf{The design sensitivity curves for TianQin, LISA, and Taiji} The sensitivity is expressed as $ h_{\mathrm{c}} $, the dimensionless characteristic strain. The black dots correspond to M31 UCXB-1 with different masses of BH.}
    \label{fig:6}
\end{figure}

    The strong GW emission of M31 UCXB-1 makes it a potential extragalactic low-frequency GW source. We employ the GW characteristic strain ($ h_{\mathrm{c}} $) to exhibit the detecbility of M31 UCXB-1 in Figure \ref{fig:6}. The sensitivity curves of LISA and Taiji are from Liu et al.,\upcitep{Liu2023} while that of TianQin is from Wang et al.\upcitep{Wang2019} Meanwhile, the GW characteristic strain for BH--WD system is from Chen.\upcitep{Chen2020} The results exhibit that M31 UCXB-1 is unlikely to be detectable by TianQin, but likely detectable by Taiji, and possibly detectable by LISA if $ M_{1} \gtrsim 10 \: M_{\odot} $.

    \subsection*{Spectrum fitting results}
    \addcontentsline{toc}{subsection}{Spectrum fitting results}
    \phantomsection
    \label{Spectrum fitting results}
    We complemented the spectrum fitting result in previous work\upcitep{Zhang2024} (\textit{Chandra} $0.5$–$8.0\ \mathrm{keV}$) with \textit{XMM--Newton} observation, extending the energy coverage while remaining within a comparable timescale. Spectral fits were performed for both the high- and low-accretion states (see Table~\ref{tab:spectra}). 
    
    Spectrum in both low and high state are adequately described by an absorbed multicolor disk blackbody plus power-law model. In the low state, the spectrum is dominated by the \texttt{powerlaw} component, with an approximate photon index ($\Gamma$). In the high state, the spectrum softens markedly: the power-law photon index steepens to $\Gamma \approx 3$, and the \texttt{diskbb} component’s share of the total 0.5–10 keV luminosity rises from 5.2 \% to 22.4 \%, reflecting a substantially stronger thermal–disk contribution. 
    Introducing an additional blackbody radiation component (\texttt{bbodyrad}) does not significantly enhance the fit, suggesting that it does not have a hard surface. In three‐component fit (\texttt{TBabs*(powerlaw+diskbb+bbodyrad)}), the \texttt{bbodyrad} temperature and normalization are effectively unconstrained, indicating that the soft component is not statistically required. Consequently, no additional soft blackbody is justified. The lack of any well‐constrained boundary‐layer emission further favors a BH accretor rather than a super‐Eddington NS interpretation.\upcitep{Lin2007}

    Our spectral analysis indicates that the source is consistent with being a BHXRB.\upcitep{Remillard2006} In the high state, the spectrum is characterized by a strong disk component originating from an optically thick accretion disk, modeled effectively with a $kT_{\mathrm{in}} \approx 2.3$ keV, accompanied by a significant steep power-law tail (photon index $\Gamma \approx 3.0$). The inner disk temperature $kT_{\mathrm{in}} \approx 2.3$ keV is relatively high for BH systems, while for most stellar-mass BHs, the typical inner disk temperature is $ kT_{\mathrm{in}} \approx 1 $ keV.\upcitep{Miller2007} Based on the relation $ f_{\mathrm{bol}} \propto r_{\mathrm{in}}^{2} T_{\mathrm{in}}^{4} D^{-2} \cdot \mathrm{cos} \: i $ ($ f_{\mathrm{bol}} $: bolometric flux, $ D $: distance),\upcitep{Kubota1998} high inclination angle $ i $ or small inner disk radius $ r_{\mathrm{in}} $ will lead to a higher $ T_{\mathrm{in}} $. Therefore, the inner disk temperature can reach to 2 keV for the extreme Kerr BH, such as GRS 1915+105 ($ i \approx 66^\circ $, $ M_{\mathrm{BH}}\approx 14 \: M_{\odot} $, $ R_{\mathrm{in}} \approx 21 $ km, spin parameter $ a_{\ast} \approx 0.98 $).\upcitep{McClintock2006} The inclination angle $ i $ of M31 UCXB-1 is difficult to constrain through observations, but its periodicity eclipsing signal disappears when it is in the low/hard state, which implies that it is a high inclination system ($ i > 80^\circ $).\upcitep{Fiocchi2011} From $ f_{\mathrm{bol}} \propto r_{\mathrm{in}}^{2} T_{\mathrm{in}}^{4} D^{-2} \cdot \mathrm{cos} \: i $, the extreme Kerr BH ($ a_{\ast}=1 $) corresponds to the lower limit of $ i $ because of the the small $ r_{\mathrm{in}} $. For M31 UCXB-1, $ M_{\mathrm{BH}} = 3 \: M_{\odot} $ corresponds to $ i \gtrsim 73.5^\circ $; while $ M_{\mathrm{BH}} = 20 \: M_{\odot} $ corresponds to $ i \gtrsim 86.7^\circ $.
    
    The low state spectrum is dominated primarily by a harder power-law component due to Comptonization within a hot and optically thin corona, with negligible thermal disk emission. Additionally, the 0.5--10.0 keV unabsorbed luminosity increases from $7.6^{+0.1}_{-0.1}	\times10^{37} \: \mathrm{erg}\: \mathrm{s}^{-1}$ in the low state to $2.9^{+0.2}_{-0.2}	\times 10^{38} \: \mathrm{erg} \: \mathrm{s}^{-1}$ in the high state. This substantial rise further indicates that the source enters the high state, in accord with the behavior of BH accretors. In contrast, NS accretors, especially under super-Eddington accretion conditions, are expected to display additional spectral features. The presence of a solid surface leads to the formation of a boundary layer between the accretion disk and the NS surface, emitting as a soft blackbody component (\texttt{bbodyrad}) with temperatures typically in the range of 0.5–2 keV. This component is expected to be prominent in both high and low state, contributing significantly to the overall spectrum.

\begin{figure}
    \centering
    \includegraphics[width=0.7\linewidth]{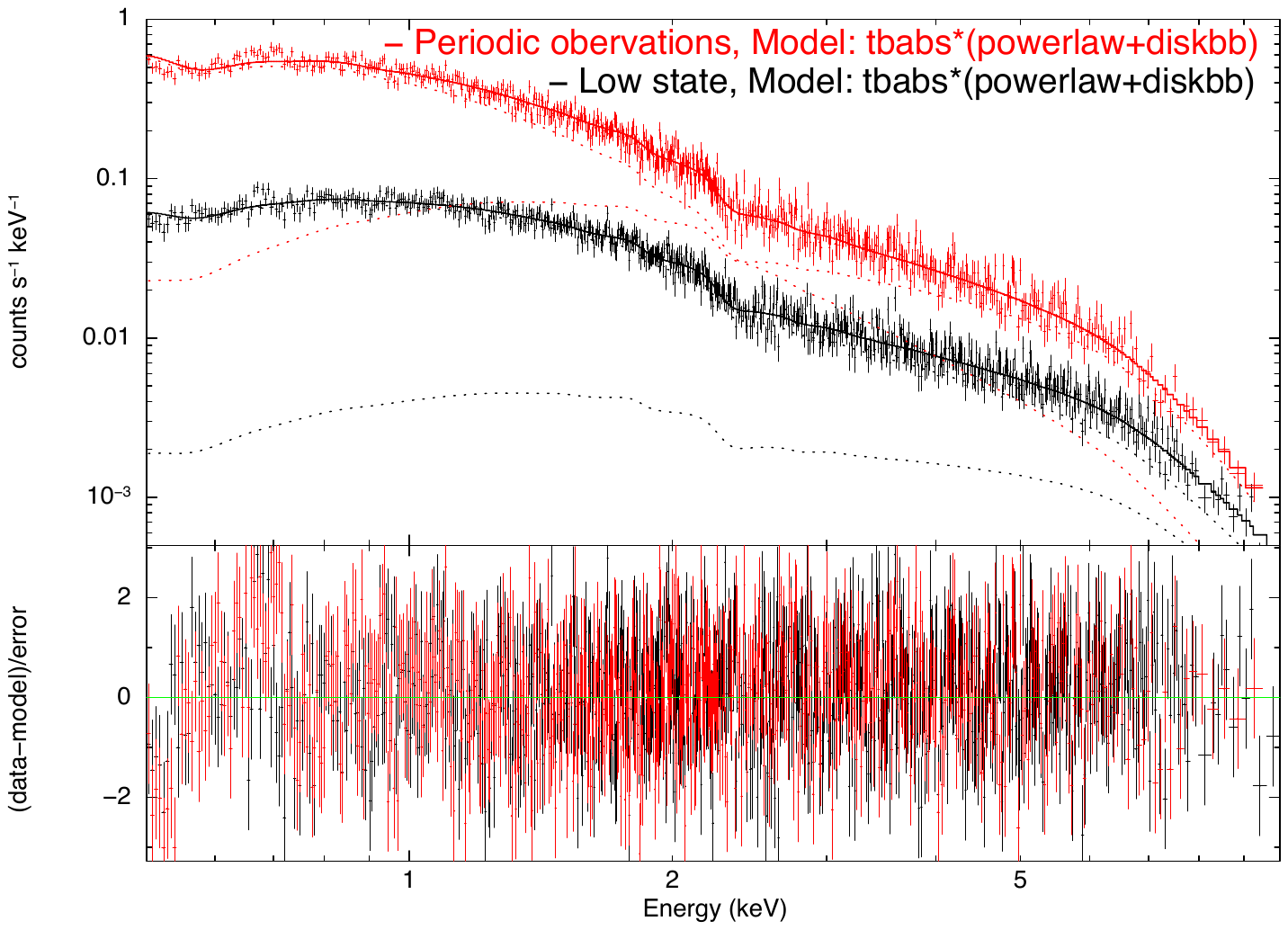}
    \captionsetup{font=normalsize, labelfont=bf}
    \caption{\textbf{\textit{XMM-Newton} spectral fitting results for periodic (red) and low-state (black) observations} Periodic-state and low state spectra are fitted with an absorbed multicolor disk + power-law model (\texttt{TBabs*(diskbb+powerlaw)}).
    The red points and curve correspond to high-luminosity state, and the black points and fit illustrate the low/hard state.}
    \label{fig:5}
\end{figure}
\begin{table*}
\centering
\caption{Spectral–fit parameters for the trial models}
\label{tab:spectra}
\resizebox{\textwidth}{!}{
\begin{tabular}{ccccc}
\hline\hline
Parameter & Low state & Low state & High state  & High state \\
Model 
  & \makecell{TBabs*(powerlaw\\+diskbb)} 
  & \makecell{TBabs*(powerlaw\\+diskbb+bbodyrad)} 
  & \makecell{TBabs*(powerlaw\\+diskbb)}  
  & \makecell{TBabs*(powerlaw\\+diskbb+bbodyrad)} \\
\hline
$N_{\mathrm{H}}$ ($10^{22}\,\mathrm{cm}^{-2}$)      
           & $0.12^{+0.02}_{-0.02}$ 
            & $0.04^{+0.08}_{-0.04}$ 
           & $0.21^{+0.03}_{-0.03}$ 
           & $0.21^{+0.03}_{-0.03}$ \\
$\Gamma$   
           & $1.5^{+0.1}_{-0.1}$ 
           & $1.1^{+0.6}_{-1.6}$
           & $3.0^{+0.3}_{-0.3}$ 
           & $3.0^{+0.3}_{-0.3}$ \\
Power--law norm ($10^{-4}$)    
           & $1.2^{+0.2}_{-0.2}$ 
           & $0.5^{+0.8}_{-0.4}$
           & $8.3^{+0.7}_{-0.6}$ 
           & $8.3^{+0.7}_{-0.6}$ \\
Luminosity fraction (Power--law)
& 94.8 \%
& 70.5 \%
& 63.8 \%
& 63.8 \% \\
$kT_{\mathrm{in}}$ (keV)$^{a}$   
           & $0.5^{+0.1}_{-0.2}$
           & $1.3^{+0.7}_{-1.3}$
           & $2.3^{+0.2}_{-0.2}$ 
           & $2.3^{+0.2}_{-0.2}$ \\
Diskbb norm                 
           & $ 0.05_{-0.03}^{+0.30} $
           & $4_{-4}^{+100}\times10^{-3}$
           & $2.6^{+1.1}_{-0.9}\times10^{-3}$ 
           & $2.6^{+1.1}_{-0.9}\times10^{-3}$ \\
Luminosity fraction (Diskbb)
& 5.2 \%
& 22.4 \%
& 36.2 \%
& 36.1 \%
\\

$kT_{\mathrm{bb}}$ (keV)$^{b}$   
           & ---
           & $0.3^{+0.05}_{-0.3}$
           & ---
           & $1^{+200}_{-1}$ \\
Bbodyrad norm               
           & ---
           & $0.9_{-0.9}^{+26}$
           & ---
           & $2_{-2}^{+20}\times10^{-4}$ \\
Luminosity Fraction (bbodyrad)
& ---
& 7.1 \%
& ---
& 0.1 \%\\
\hline
$\chi^{2}$/d.o.f & $704/728$ & $702/726$ & $728/628$ & $728/626$ \\
\hline
\end{tabular}}
\begin{flushleft}
\textbf{Notes.} Errors are 90\,\% confidence for one parameter of interest
($\Delta \chi^{2}=2.706$).  
$(a)$ Inner–disc temperature from \texttt{diskbb}.
$(b)$ Blackbody temperature from \texttt{bbodyrad}.  
\end{flushleft}
\end{table*}

It is also a UCXB in the high state, providing direct evidence that UCXBs hosting BH accretors can reach the same extreme accretion regimes and disk–corona configurations characteristic of canonical BH X-ray binaries. This finding offers new constraints on accretion physics and on identifying the accretors in UCXBs.


	\section*{DISCUSSION} \label{DISCUSSION}
    \addcontentsline{toc}{section}{DISCUSSION}
    \phantomsection
    \label{DISCUSSION}

\begin{figure}
    \centering
    \captionsetup{font=normalsize, labelfont=bf}
    \includegraphics[width=0.6\linewidth]{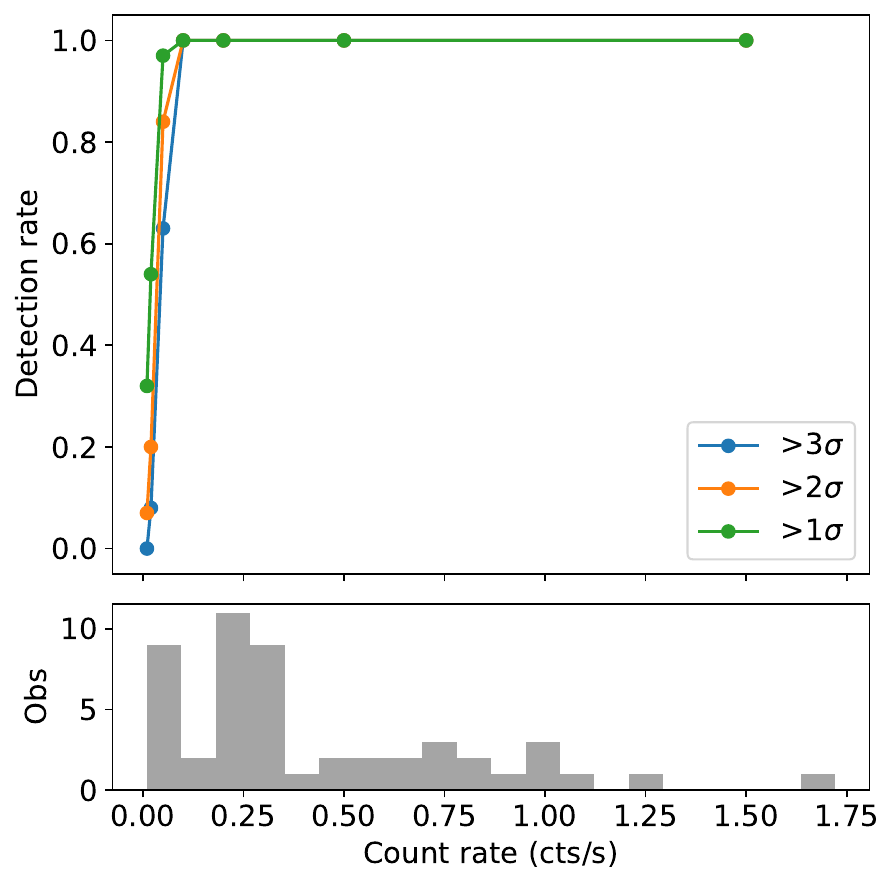}
    \caption{\textbf{Results of the injection–recovery simulations for eclipse signals} Artificial light curves were generated with an eclipse fraction of 0.3 in phase, an eclipse depth of 0.4, and a total duration of 20 ks, consistent with typical \textit{XMM-Newton} exposures. The detection rate is shown as a function of count rate. The high recovery fraction across most simulated datasets demonstrates that the Lomb–Scargle method is nearly complete for these parameters, ruling out methodological bias and confirming that the non-detections in real data are intrinsic rather than due to sensitivity limitations.}
    \label{fig:8}
\end{figure}

    In this work, we analyzed 50 observations from \textit{XMM-Newton} of M31 UCXB-1, with periodic signals detected in only 5 observations, which correspond to its brightest phases. To test whether this could be due to a detection bias of the Lomb-Scargle method, we perform the injection-recovery simulations using artificially generated eclipse light curves. The simulated eclipses were constructed with an eclipse fraction of 0.3 in phase, a depth of 0.4, and a total light curve duration of 20 ks,\upcitep{Bao2020} consistent with the typical exposure times of most observations. We find that for the majority of the simulated light curves, the Lomb-Scargle method is able to recover the injected signal with high completeness (see Figure~\ref{fig:8}). This rules out the possibility that the non-detections are caused by limitations of period-searching methods. Instead, the results indicate that the periodic signal is only present during the high luminosity states of the source, suggesting that the appearance of the eclipse is linked to the accretion state and the visibility of the compact object's orbital modulation. Moreover, in the observations where no periodicity was detected, the Lomb–Scargle power around 465.32 s remains below the $2\sigma$ level, further supporting the absence of a coherent periodic signal in those data.

\begin{figure}
    \centering
    \captionsetup{font=normalsize, labelfont=bf}
  \begin{overpic}[width=0.49\linewidth]{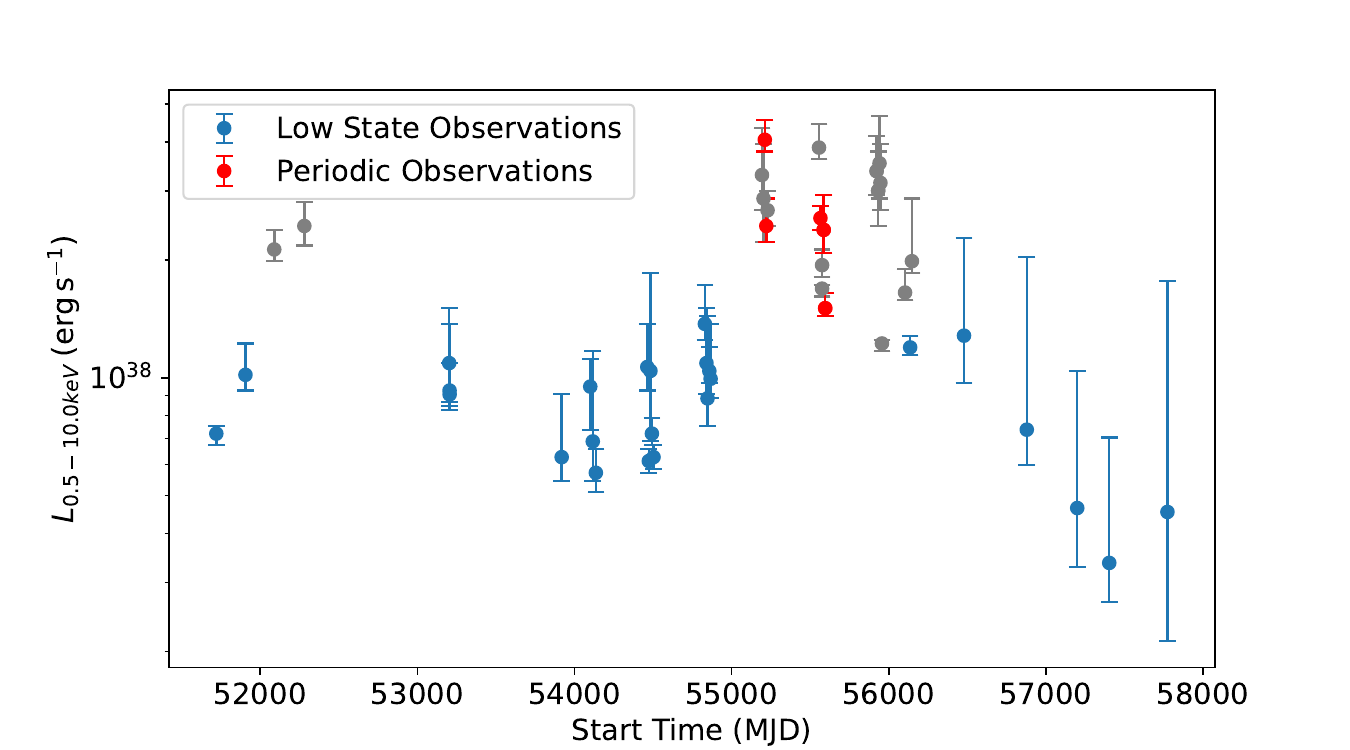}
    \put(2,45){\small A}
  \end{overpic}
  \begin{overpic}[width=0.49\linewidth]{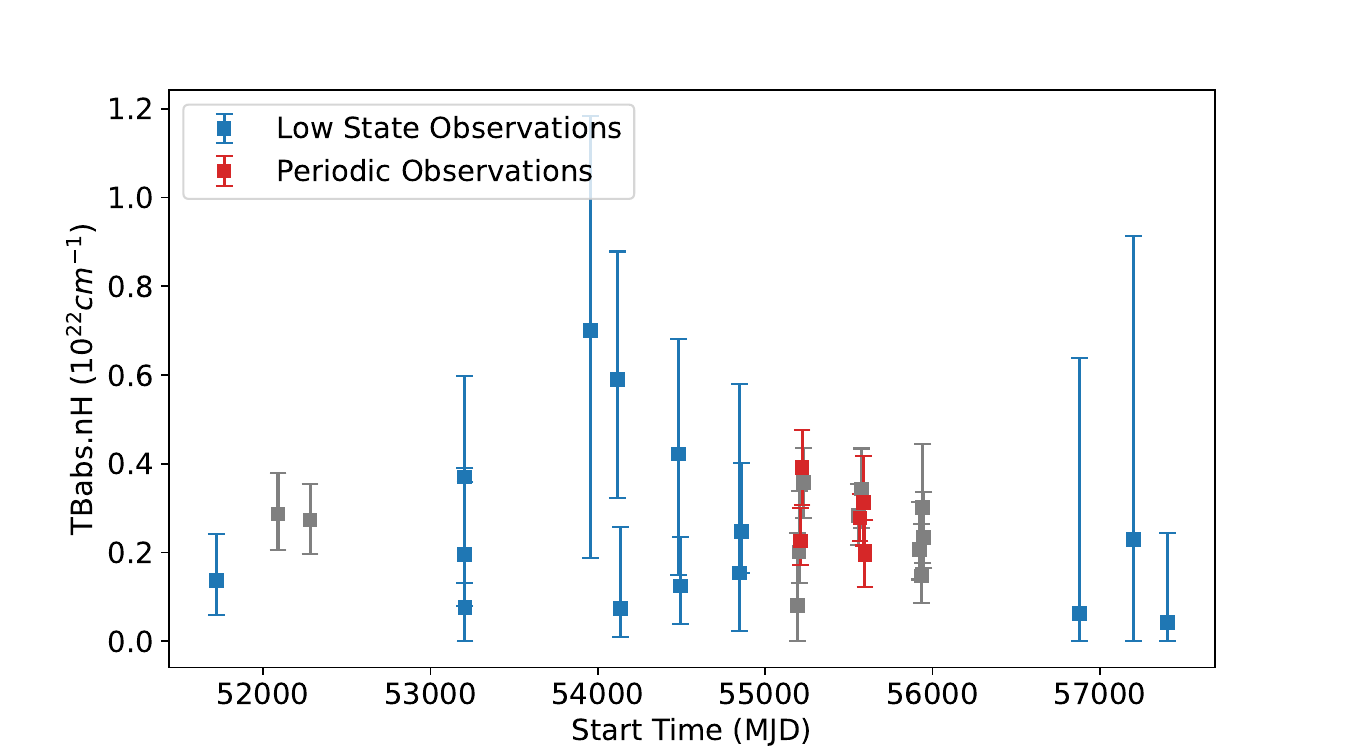}
    \put(2,45){\small B}
  \end{overpic}

  \begin{overpic}[width=0.49\linewidth]{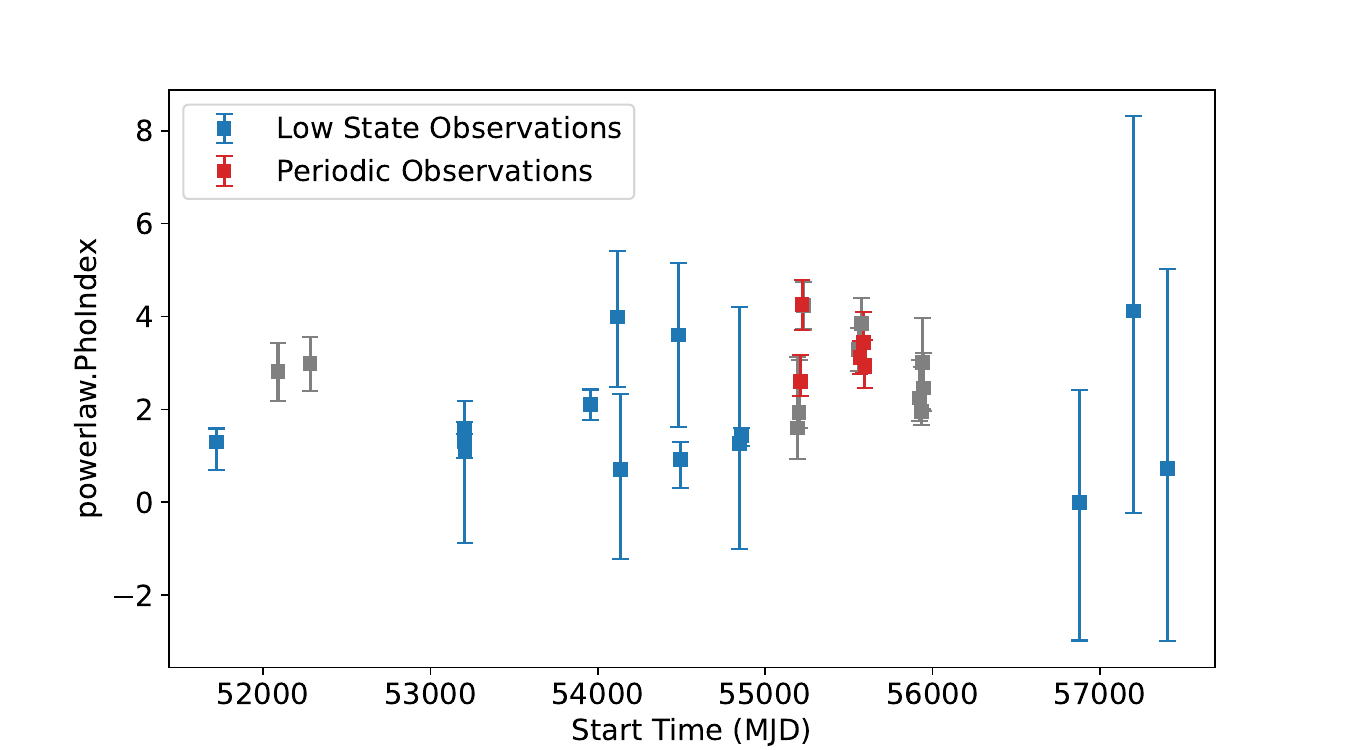}
    \put(2,45){\small C}
  \end{overpic}
  \begin{overpic}[width=0.49\linewidth]{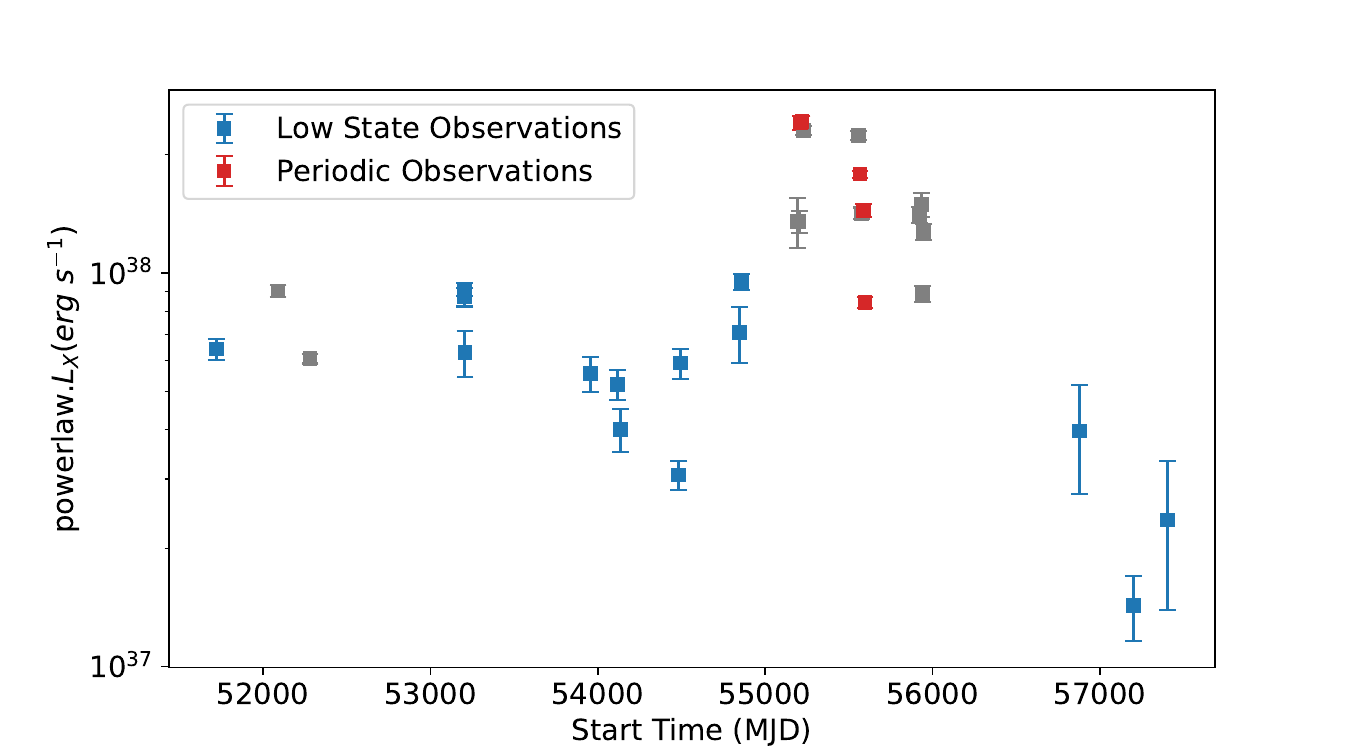}
    \put(2,45){\small D}
  \end{overpic}

  \begin{overpic}[width=0.49\linewidth]{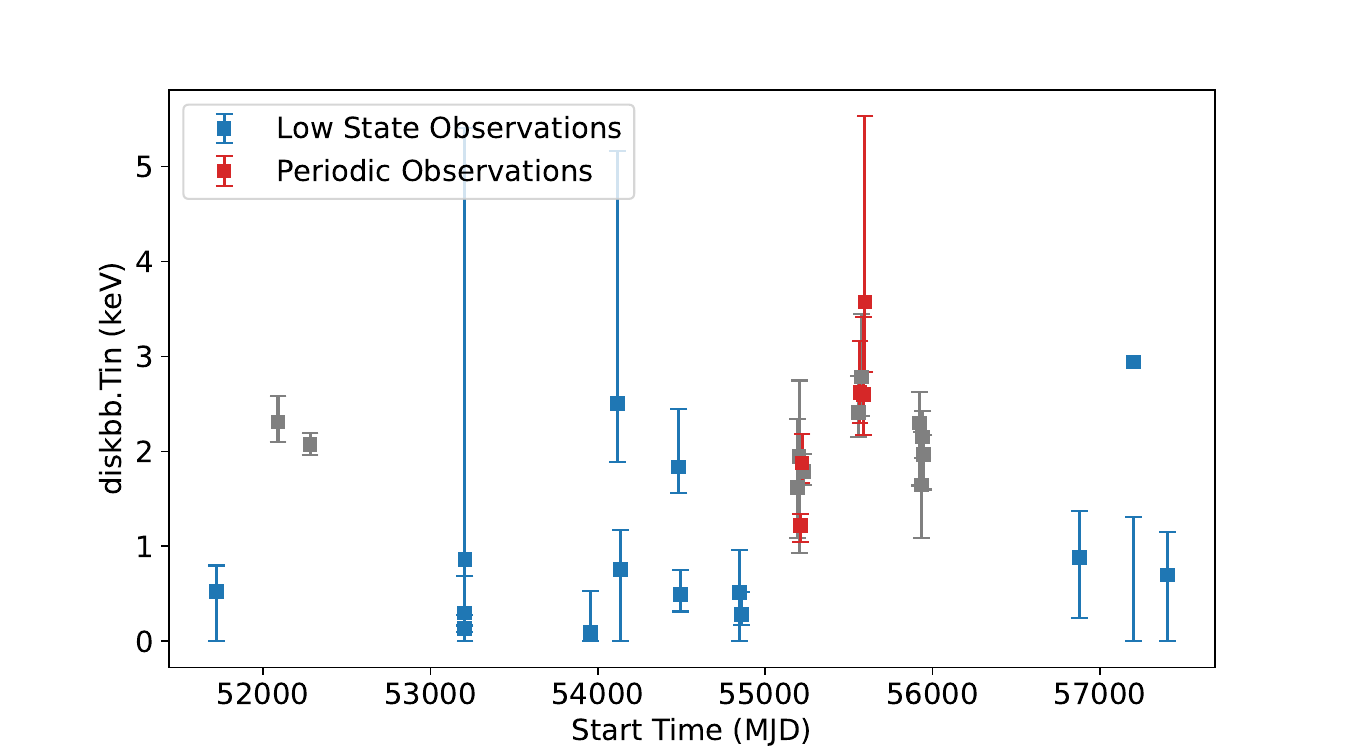}
    \put(2,45){\small E}
  \end{overpic}
  \begin{overpic}[width=0.49\linewidth]{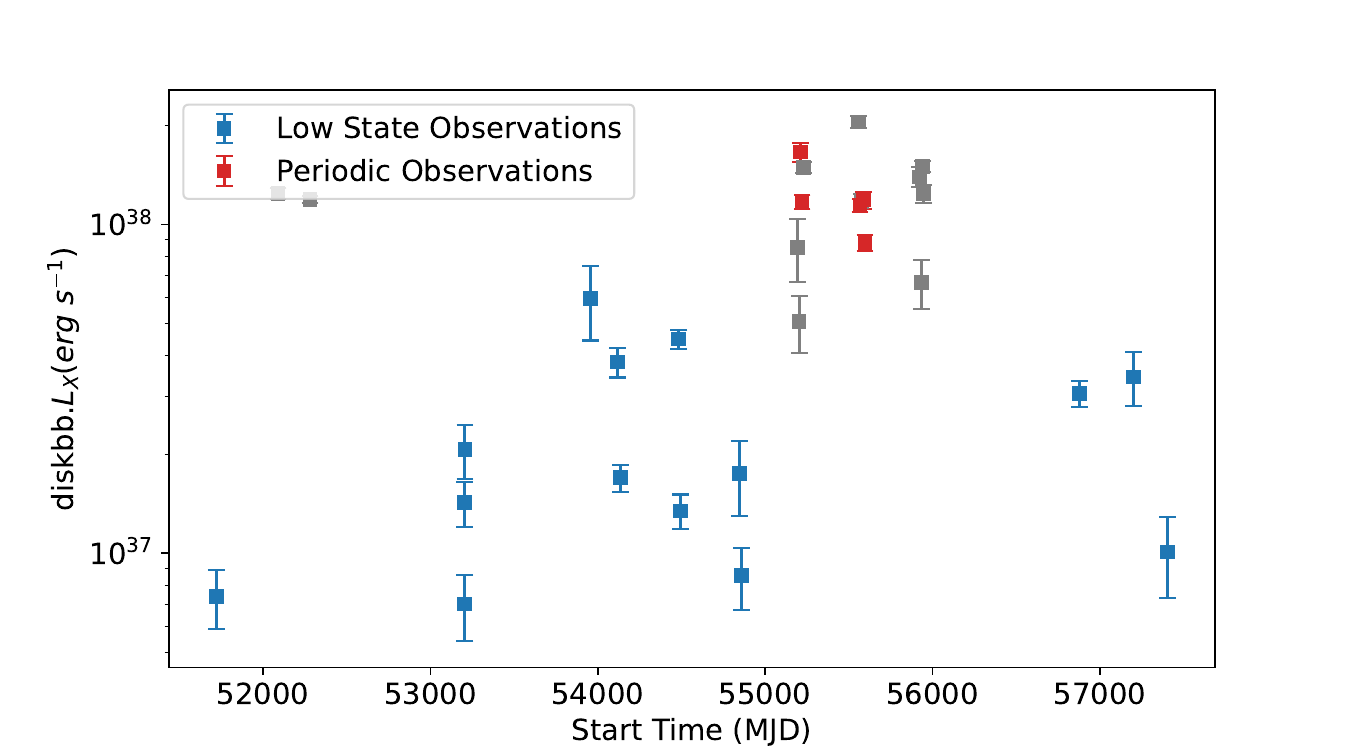}
    \put(2,45){\small F}
  \end{overpic}
    \caption{
    \textbf{Temporal evolution of spectral fit parameters for \textit{XMM-Newton} observations} All observations are modeled with \texttt{TBabs*(powerlaw+diskbb)} and plotted at its Modified Julian Date. The six panels are arranged as follows:
(A) Total unabsorbed 0.5–10 keV luminosity.
(B) Hydrogen column density $N_{\mathrm{H}}$.
(C) Power-law photon index $\Gamma$.
(D) Unabsorbed 0.5–10 keV luminosity of the power-law component.
(E) Inner disk temperature $kT_{\mathrm{in}}$ of the disk-blackbody.
(F) Unabsorbed 0.5–10 keV luminosity of the disk-blackbody component.
Error bars denote 90 \% confidence intervals.}
    \label{fig:7}
\end{figure}
    
        Other UCXBs also exhibit similar phenomena, for instance, the periodic signal disappears in the low/hard state of 4U 0513--40,\upcitep{Fiocchi2011} and we speculate that M31 UCXB-1 also has the similar physical mechanism. During the high/soft state, the soft thermal emission comes from a small region near the compact object, which becomes the origin of the periodic modulation. However, when it transfers to the low/hard state, the soft thermal emission area disappears and the hot corona dominates the radiation. The corona is often an extended structure, larger than the soft X-ray emission region, which is difficult to be obscured for producing periodic signals.\upcitep{Fiocchi2011}
        
        The results of the theoretical analyses demonstrate that M31 UCXB-1 is a BH/NS--WD UCXB with the shortest orbital period. Most UCXBs follow the predicted $ L_{\mathrm{2-10 \: keV}} - P_{\mathrm{orb}} $ relationship, where shorter orbital periods correspond to higher luminosities, as shown in Figure \ref{fig:3}. However, the theoretical model can only distinguish BH and NS systems when $ P_{\mathrm{orb}} > 1000 \: \mathrm{s} $. The reason for this is that as $ P_{\mathrm{orb}} $ increases, $ \dot{M}_{1} $ decreases, the energy conversion efficiency for NS remains approximately constant due to their hard surfaces, but the energy conversion efficiency for BH declines with reduced $ \dot{M}_{1} $ (see the \nameref{BH/NS--WD binary system} section).\upcitep{Narayan1995, Watarai2001} Among all the known UCXBs (black dots) exhibited in Figure \ref{fig:3}, only 47 Tuc X-9 is identified as a BH candidate due to its high radio-to-X-ray flux ratio,\upcitep{Miller-Jones2015} which agrees well with the theoretically predicted $ L_{\mathrm{2-10 \: keV}}-P_{\mathrm{orb}} $ relation.
        
        It should be noted that the X-ray luminosities for 2S 0918--549, 4U 1543--624, and IGR J17062--6143 in Figure \ref{fig:3} show significant deviations from the theoretical predictions. We employ $ L_{\mathrm{2-10 \: keV}} = 4 \pi D^{2} F_{\mathrm{2-10 \: keV}} $ to estimate the X-ray luminosities for the known UCXBs, where $ D $ is the distance and $ F_{\mathrm{2-10 \: keV}} $ is the flux in 2--10 keV. Therefore, the uncertainty of $ D $ will lead to significant deviations in $ L_{\mathrm{2-10 \: keV}} $. The distance estimations of NSs are primarily derived from two methods. The one is directly obtained from the distances of their host globular clusters, which is relatively reliable; while the other is from the peak flux of their type I X-ray bursts, but it has significant errors.\upcitep{Armas Padilla2023} In Figure \ref{fig:3}, the distances for 2S 0918--549, 4U 1543--624, and IGR J17062--6143 are all estimated by the helium type I X-ray bursts, while the others are derived from their host globular clusters.\upcitep{Armas Padilla2023} Apart from these three sources, for white dwarf UCXBs with accurately measured distances in Figure \ref{fig:3}, the predicted luminosity is relatively reliable. Therefore, the $ L_{\mathrm{X}}-P_{\mathrm{orb}} $ diagram holds potential value for the identification of white dwarf UCXBs.

        The $ M $-$ R $ relation for WD we employed is from Verbunt et al.\upcitep{Verbunt1988} (see Equation (\ref{eq:3}) in the \nameref{BH/NS--WD binary system} section), and following this we obtain that $ M_{2} = 0.088 \: M_{\odot} $ and $ R_{\mathrm{WD}} = 0.025 \: R_{\odot} $. This is the most commonly used $ M $-$ R $ relation for the WD, which is derived under the assumption of no Coulomb interaction,\upcitep{Nauenberg1972} and applicable to helium (He), carbon (C), and oxygen (O) WDs (electron mean molecular weight equal to 2). Deloye et al.\upcitep{Deloye2003} considered Coulomb interaction, and provided a more complicated $M$-$R$ relation. Following this, the results are only slightly changed, i.e., $ M_{\mathrm{WD}} = 0.097 \: M_{\odot} $, $ R_{\mathrm{WD}} = 0.027 \: R_{\odot} $ for the He--WD case; and $ M_{\mathrm{WD}} = $$ 0.087 \: M_{\odot} $/$ 0.083 \: M_{\odot} $ (pure C/pure O), $ R_{\mathrm{WD}} = 0.026 \: R_{\odot}$/$ 0.025 \: R_{\odot} $ (pure C/pure O) for the C/O--WD case. Therefore, the influence of elemental composition and Coulomb interaction on $ M_{\mathrm{WD}} $ and $ R_{\mathrm{WD}} $ is not significant.

        We cannot fully rule out an NS as the primary in M31 UCXB-1, as some NSs also lack the surface emissions. The possible reasons for that include (i) absorption (high inclination system): when the surface emission of NS is absorbed by the outer regions of the accretion disk or the outflow by the companion star, the blackbody radiation component will decrease significantly, such as X 1624--490;\upcitep{Iaria2007} (ii) strong Comptonization: NS is surrounded by the optically thick gas, where strong Comptonization obscures the radiation originating from its surface, such as 4U 1608--52;\upcitep{Gierliński2002} (iii) a heavy NS: the heavy NS generally has smaller radius,\upcitep{Lattimer2001} and the strong redshift on its surface will cause the observed temperature lower than the real temperature,\upcitep{Lattimer2004} which significantly reduces the contribution of its surface emission. The lack of blackbody component in M31 UCXB-1 is not likely attributable to (i) or (ii), since the blackbody deficit caused by (i) and (ii) is temporary,\upcitep{Iaria2007, Gierliński2002} whereas in M31 UCXB-1 the blackbody component is not observed in any state. Therefore, if M31 UCXB-1 harbors an NS as the accretor, (iii) is the more likely explanation for the exclusion of the surface emission: a heavy and small NS hides the surface emission, making the blackbody component unobservable in any state.
	\section*{METHODS} \label{METHODS}
    \addcontentsline{toc}{section}{METHODS}
    \phantomsection
    \label{METHODS}
    
	\subsection*{BH/NS--WD binary system}
    \addcontentsline{toc}{subsection}{METHODS}
    \phantomsection   
    \label{BH/NS--WD binary system}
        We assume that the eccentricity of the BH/NS--WD system is $ e = 0 $, and the accretion occurs when the WD fills its Roche lobe. The Roche lobe radius of the companion star is\upcitep{Paczynski1971}
        \begin{equation}
            R_{\mathrm{L2}} = 0.462 \left( \frac{M_{2}}{M} \right)^{\frac{1}{3}} a, \label{eq:1}
        \end{equation}
        where $ M = M_{1} + M_{2} $, and $ a $ is the orbital separation. Combining Equation (\ref{eq:1}) with $ R_{\mathrm{L2}} = R_{\mathrm{WD}} $ (the WD fills its Roche lobe) and $ GM / a^{3} = 4 \pi^{2} / P_{\mathrm{orb}}^{2} $ (Kepler's third law), the following relation is obtained:\upcitep{Frank2002}
        \begin{equation}
            P_{\mathrm{orb}} = \left( \frac{4 \pi^{2}}{0.462^{3}} \right)^{\frac{1}{2}} \frac{R_{\mathrm{WD}}^{\frac{3}{2}}}{(G M_{2})^{\frac{1}{2}}}, \label{eq:2}
        \end{equation}
        and $ R_{\mathrm{WD}} $ is expressed as a function of $ M_{2} $ (WD's $M$-$R$ relation):\upcitep{Verbunt1988}
        \begin{equation}
            R_{\mathrm{WD}} = g(M_{2}) = 0.0114 \: R_{\odot} \:[(\frac{M_{2}}{M_{\mathrm{Ch}}})^{-\frac{2}{3}} - (\frac{M_{2}}{M_{\mathrm{Ch}}})^{\frac{2}{3}}]^{\frac{1}{2}} \times [1 + 3.5(\frac{M_{2}}{M_{\mathrm{p}}})^{-\frac{2}{3}} + (\frac{M_{2}}{M_{\mathrm{p}}})^{-1}]^{-\frac{2}{3}}, \label{eq:3}
        \end{equation}
        where $ M_{\mathrm{Ch}} =1.44 \: M_{\odot} $ and $ M_{\mathrm{p}} = 0.00057 \: M_{\odot} $. By substituting $ g(M_{2}) $ into Equation (\ref{eq:2}), $ P_{\mathrm{orb}} $ can be expressed as a function of $ M_{2} $: $ P_{\mathrm{orb}} \propto M_{2}^{-1/2} \cdot [g(M_{2})]^{3/2} $. Therefore, $ P_{\mathrm{orb}} $ depends solely on $ M_{2} $, and once $ P_{\mathrm{orb}} $ is derived from the observations, a unique $ M_{2} $ can be determined.
        
        Under the combined influence of GW radiation and mass transfer, the expanding rate of the WD and its Roche lobe are equal ($ \dot{R}_{\mathrm{L2}} = \dot{R}_{\mathrm{WD}} $), resulting in stable accretion. Based on this, we can derive the accretion rate of the BH/NS--WD system:\upcitep{Dong2018}
        \begin{equation}
			\dot{M}_{1} = -\dot{M}_{2} = \frac{64 G^{3} M_{1} M_{2}^{2} M}{5 c^{5} a^{4} (\frac{5}{3} - \frac{\delta}{3} - 2q) }, \label{eq:4}
		\end{equation}
		where $ q = M_{2} / M_{1} $, and
        \begin{equation}
            \delta = \frac{ (\frac{M_{2}}{M_{\mathrm{Ch}}})^{-\frac{2}{3}} +  (\frac{M_{2}}{M_{\mathrm{Ch}}})^{\frac{2}{3}}}
		{ (\frac{M_{2}}{M_{\mathrm{Ch}}})^{-\frac{2}{3}} -  (\frac{M_{2}}{M_{\mathrm{Ch}}})^{\frac{2}{3}} } - \frac
		{ 2 \frac{ M_{\mathrm{p}} }{ M_{2} } [\frac{7}{3} (\frac{ M_{\mathrm{p}} }{ M_{2} })^{-\frac{1}{3}} + 1] }
		{ 1 + 3.5(\frac{ M_{2} }{ M_{\mathrm{p}} })^{ -\frac{2}{3} } + (\frac{M_{2}}{M_{\mathrm{p}}})^{-1} }. \label{eq:5}
        \end{equation}

		The Eddington luminosity for NS is $ L_{\mathrm{Edd, \: NS}} = 4 \pi G M_{\mathrm{NS}} c / \kappa_{\mathrm{es}} $, and for BH is $ L_{\mathrm{Edd, \: BH}} = 4 \pi G M_{\mathrm{BH}} c / \kappa_{\mathrm{es}} $, where the electron scattering opacity is $ \kappa_{\mathrm{es}} = 0.2 \: \mathrm{cm}^{2} \: \mathrm{g}^{-1} $ (electron mean molecular weight equal to 2). Therefore we can define the Eddington accretion rates for NS and BH, respectively: $ \dot{M}_{\mathrm{Edd, \: NS}} = L_{\mathrm{Edd, \: NS}}/ (0.2 c^{2}) $ and $ \dot{M}_{\mathrm{Edd, \: BH}} = L_{\mathrm{Edd, \: BH}} / (0.1 c^{2}) $, where the constants 0.2 and 0.1 are employed from Narayan et al.\upcitep{Narayan1995} In addition, we define two dimensionless parameters, $ \dot{m}_{\mathrm{NS}} = \dot{M}_{1} / \dot{M}_{\mathrm{Edd, \: NS}} $ and $ \dot{m}_{\mathrm{BH}} = \dot{M}_{1} / \dot{M}_{\mathrm{Edd, \: BH}} $. Following these hypotheses, the bolometric luminosity for NS and BH systems are respectively estimated as
		\begin{equation}
			L_{\mathrm{bol, \: NS}} =  L_{\mathrm{Edd, \: NS}} \cdot \dot{m}_{\mathrm{NS}},  \label{eq:6}
		\end{equation}
		\begin{equation}
			L_{\mathrm{bol, \: BH}} = 
			\left\{
			\begin{aligned}
				&L_{\mathrm{Edd,\: BH}} \cdot (1 + \mathrm{ln} \: \dot{m}_{\mathrm{BH}})  & \dot{m}_{\mathrm{BH}} \geq 1, \\
				&L_{\mathrm{Edd, \: BH}} \cdot \dot{m}_{\mathrm{BH}} & 0.01 \leq \dot{m}_{\mathrm{BH}} < 1, \\
				& L_{\mathrm{Edd, \: BH}} \cdot 100 \: \dot{m}_{\mathrm{BH}}^{2} &   \dot{m}_{\mathrm{BH}} \leq 0.01.
			\end{aligned}
			\right. \label{eq:7}
		\end{equation}
		For NS, the energy conversion efficiency remains $ \eta = 0.2 $, and therefore the luminosity remains $ L_{\mathrm{bol, \: NS}} \propto \dot{m}_{\mathrm{NS}} $, regardless of the accretion rate.\upcitep{Narayan1995} For BH, we assume that the accretion state changes when $ \dot{m}_{\mathrm{BH}} = 0.01 $ and $ \dot{m}_{\mathrm{BH}} = 1 $. $ \dot{m}_{\mathrm{BH}} \leq 0.01 $ corresponds to the advection-dominated accretion flow (ADAF), in this condition the energy conversion efficiency is $ \eta \propto \dot{m} $,\upcitep{Esin1997, Narayan1998} and thus $ L_{\mathrm{bol, \: BH}} \propto \dot{m}_{\mathrm{BH}}^{2} $. $ 0.01 \leq \dot{m}_{\mathrm{BH}} < 1 $ corresponds to the standard thin disk, in this condition $ L_{\mathrm{BH}} \propto \dot{m}_{\mathrm{BH}} $.\upcitep{Esin1997, Narayan1998} $ \dot{m}_{\mathrm{BH}} \geq 1 $ corresponds to the slim accretion disk, in this condition $ L_{\mathrm{bol, \: BH}} \propto 1 + \mathrm{ln} \: \dot{m}_{\mathrm{BH}} $.\upcitep{Abramowicz1988, Watarai2001} The luminosity in 2--10 keV is expressed as 
       \begin{equation}
           L_{\mathrm{2-10 \: keV}} = \lambda \: L_{\mathrm{bol}}, \label{eq:8}
       \end{equation} 
       where $ \lambda $ is the thermal correction factor at 2--10 keV, and we employ $ \lambda = 0.1-0.5 $ for both BH and NS systems.\upcitep{Anastasopoulou2022} The width of the orange and green bands in Figure \ref{fig:3} is determined by the range of $ \lambda $, with the lower boundary corresponding to $ \lambda = 0.1 $ and the upper boundary corresponding to $ \lambda = 0.5 $. In addition, the observed 2--10 keV luminosity of those known UCXBs are derived by $ L_{\mathrm{2-10 \: keV}} = F_{\mathrm{2-10 \: keV}} \cdot 4 \pi D^{2} $, where $ F_{\mathrm{2-10 \: keV}} $ is the X-ray flux and $ D $ is the distance. The observed $ F_{\mathrm{2-10 \: keV}} $ and $ D $ are from Armas Padilla et al.\upcitep{Armas Padilla2023}
       
\subsection*{Data analysis procedure} 
\addcontentsline{toc}{subsection}{Data analysis procedure}
\phantomsection
\label{Data analysis procedure}

Because \textit{XMM-Newton} registers a higher count rate than \textit{Chandra}, we carried out a search for coherent modulations in all 50 available \textit{XMM-Newton} observations by Lomb-Scargle method. Only five data sets (ObsIDs 0600660401, 0600660501, 0650560301, 0650560501, 0650560601) exhibit a statistically significant periodic signal. By merging the photon arrival times from these five observations, we refine the period to $ P=465.32 $ s.

We use the observation data targeted at M31 from 2003 to 2017 from \textit{XMM-Newton} for spectrum fitting. All EPIC data were reprocessed with SAS v22.1.0 and the matching CCF set. Photon times were converted to Barycentric Dynamical Time with \texttt{barycen}. 10–12 keV single-pixel light curves then defined GTIs ($\mathrm{\leq 0.3 \ cts\ s^{-1}}$ for MOS, $\mathrm{\leq 0.6 \ cts\ s^{-1}}$ for PN) to excise soft-proton flares in the same TDB frame. GTI-filtered events provided barycentre-corrected lists from which source and background spectra were extracted within a  $20 ''$ circle region and calibrated with \texttt{backscale}, \texttt{rmfgen}, and \texttt{arfgen}, yielding final spectra ready for joint fitting. We restrict our spectral analysis to the 0.5–10.0 keV energy band, to minimize the effect from soft-proton flares and particle background.

To investigate the spectral behavior in different luminosity regimes, we constructed two composite spectra. The first combines 29 observations in a persistently low state, while the second stacks the five periodic (high state) observations listed above. Observations shown in grey in Figure \ref{fig:7} fall into an intermediate, luminosity-varying state. They are excluded from the joint spectral fits because their mixed flux levels would bias the results. Low‐state observations are shown in blue in Figure~\ref{fig:7}, while periodic observations are shown in red. We fitted the combined spectra using \texttt{XSPEC} v12.12.1. Each dataset was modeled with \texttt{TBabs*(powerlaw+diskbb)} and \texttt{TBabs* (powerlaw+diskbb+bbodyrad)}. The best‐fit parameters are listed in Table~\ref{tab:spectra}, and the spectra with the \texttt{TBabs*(powerlaw+diskbb)} model are shown in Figure~\ref{fig:5}. We also fitted the individual spectra from 29 observations (each with sufficient photon counts) using the \texttt{TBabs*(powerlaw+diskbb)} model. The evolution of the best-fit parameters over time is shown in Figure~\ref{fig:7}.

    \section*{DATA AND CODE AVAILABILITY}
    The data is available and can be accessed online at \url{https://nxsa.esac.esa.int/nxsa-web/}. Any code used in this study is available from the corresponding author upon reasonable request.

	\section*{ACKNOWLEDGMENTS}
	This work was supported by the National Key R\&D Program of China under grants 2023YFA1607901 and 2021YFA1600401, the National Natural Science Foundation of China under grants 12433007, 12221003, 12225302, and 12473041. We also acknowledge the science research grants from the China Manned Space Project with No. CMS-CSST-2025-A13. The funders had no role in study design, data collection and analysis, decision to publish, or preparation of the manuscript.

    \section*{AUTHOR CONTRIBUTIONS}
    W.-M.G. and Z.L. proposed the project, and the manuscript was written by Q.-Q.M. and J.Z. All authors contributed to the manuscript.

\end{document}